# Comments on the paper: Synthesis, growth and characterization of cadmium manganese thiocyanate (CMTC) crystal

Bikshandarkoil R. Srinivasan,
Department of Chemistry, Goa University, Goa 403 206, India
Email: srini@unigoa.ac.in Telephone: 0091-(0)832-6519316; Fax: 0091-(0)832-2451184

**Abstract**

The authors of the title paper (Spectrochim. Acta 79A (2011) 340-343) report the crystal growth of cadmium manganese thiocyanate $CdMn(SCN)_4$ and its characterization by single crystal X-ray diffraction and infrared spectrum. Many points of criticism, concerning the crystal growth and reported experimental data are highlighted in this report.

**Introduction**

Rosenheim and Cohn reported on the synthesis of the first examples of bimetallic thiocyanates of general formula $AB(SCN)_4$ (A = bivalent metal; B = Hg) in 1901 [1]. Since then, these compounds have been the subject of several investigations [2-14]. This more than century long research has enriched our understanding of the synthetic aspects, spectral characteristics, structural features and material applications of the $AB(SCN)_4$ crystals and the related Se analogues. Several of the structurally characterized $AB(SCN)_4$ compounds are isostructural and crystallize in the non-centrosymmetric tetragonal $I\bar{4}$ space group (Table 1). The thiocyanate ion functions as a bridge between the bivalent (A) metal ion and the (B) metal ion in the $AB(SCN)_4$ compounds, resulting in a three-dimensional polymeric structure consisting of $\{BS_4\}$ and $\{AN_4\}$ tetrahedra (Fig. 1). A recent paper reported the growth and characterization of a so called cadmium manganese thiocyanate crystal abbreviated by the authors as CMTC [15]. Unfortunately, the results reported in [15] do not in any way concern with $CdMn(SCN)_4$ as claimed by the authors and the title paper has many points of criticism, which are highlighted herein. In order to avoid the non-standard abbreviation CMTC and also its known use for another well characterized crystal namely cadmium mercury thiocyanate $CdHg(SCN)_4$ [8], the title compound is referred to by its formula $CdMn(SCN)_4$ in this comment.

**Table 1** Geometry of metals A and B in thiocyanate bridged AB(SCN)$_4$ crystals

| No | Compound | Space group | Geometry around A | Geometry around B | Ref |
|---|---|---|---|---|---|
| 1 | CuHg(SCN)$_4$ | *Pbcn* | {CuN$_4$} square | {HgS$_4$} tetrahedron | 3 |
| 2 | CuHg(SCN)$_4$ | *C2/c* | {CuN$_4$} square | {HgS$_4$} tetrahedron | 4 |
| 3 | CoHg(SCN)$_4$ | *I$\bar{4}$* | {CoN$_4$} tetrahedron | {HgS$_4$} tetrahedron | 5 |
| 4 | ZnHg(SCN)$_4$ | *I$\bar{4}$* | {ZnN$_4$} tetrahedron | {HgS$_4$} tetrahedron | 7 |
| 5 | CdHg(SCN)$_4$ | *I$\bar{4}$* | {CdN$_4$} tetrahedron | {HgS$_4$} tetrahedron | 8 |
| 6 | MnHg(SCN)$_4$ | *I$\bar{4}$* | {MnN$_4$} tetrahedron | {HgS$_4$} tetrahedron | 6 |
| 7 | MnHg(SeCN)$_4$ | *I$\bar{4}$* | {MnN$_4$} tetrahedron | {HgSe$_4$} tetrahedron | 14 |
| 8 | CoHg(SeCN)$_4$ | *I$\bar{4}$* | {CoN$_4$} tetrahedron | {HgSe$_4$} tetrahedron | 13 |
| 9 | CdHg(SeCN)$_4$ | *I$\bar{4}$* | {CdN$_4$} tetrahedron | {HgSe$_4$} tetrahedron | 10, 12 |
| 10 | ZnCd(SCN)$_4$ | *I$\bar{4}$* | {ZnN$_4$} tetrahedron | {CdS$_4$} tetrahedron | 9 |
| 11 | ZnCd(SeCN)$_4$ | *I$\bar{4}$* | {ZnN$_4$} tetrahedron | {CdSe$_4$} tetrahedron | 11, 12 |

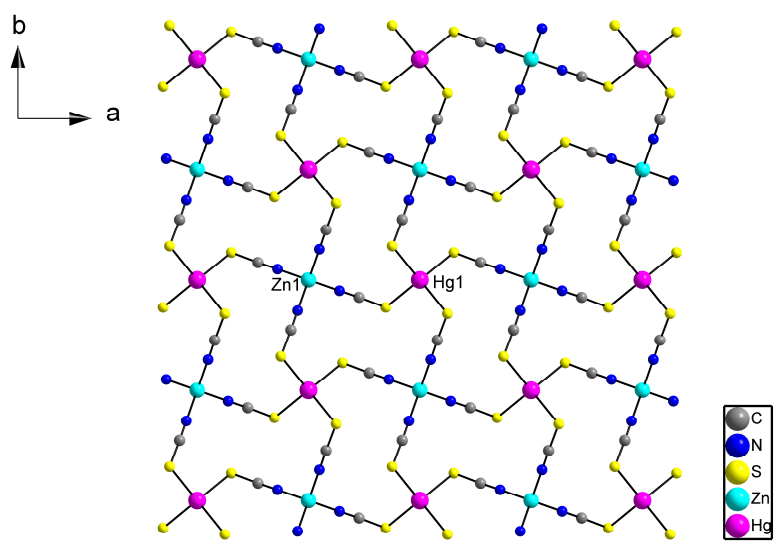

**Fig. 1.** A portion of the polymeric chain of ZnHg(SCN)$_4$ viewed along *c* axis showing the tetrahedral {HgS$_4$} and {ZnN$_4$} units (Figure is drawn by using the cif data in [7])

**Comment**

*Structural chemistry of AB(SCN)$_4$ crystals*

According to the authors of the title paper, their aim was to attempt to replace Hg in CdHg(SCN)$_4$ by Mn for growth of CdMn(SCN)$_4$ crystal [15]. Recently it has been shown that structural chemistry principles are useful to determine beforehand, which atom or ion or molecule can be incorporated into the known crystal structure of a compound [16]. In view of the availability of a large number of structurally characterized AB(SCN)$_4$ crystals, their known structural aspects can be used to predict if substitution of Hg in CdHg(SCN)$_4$ by Mn to grow crystals of CdMn(SCN)$_4$, as mentioned by the authors of the title paper, is actually feasible or not. A scrutiny of the compounds listed in Table 1 reveals that of the two metals A and B, the larger Hg(II) ion (1.10 Å) is always bonded to thiocyanate at the sulfur site (selenium of the selenocyante), while the smaller bivalent metal ions namely Co (0.72 Å), Zn (0.74 Å), Mn (0.80 Å) and Cd (0.92 Å) are bonded to the (SCN)$^-$ ion at the nitrogen end. Values in bracket are the Shannon and Prewitt ionic radii of the metal [17]. The compounds ZnCd(SCN)$_4$ and ZnCd(SeCN)$_4$ are two examples in which Cd(II) (0.92 Å) occupies the B site, while the smaller Zn(II) (0.74 Å) ion is situated in the A site. However, in CdHg(SCN)$_4$ the smaller Cd(II) ion occupies the A site and is bonded to N of (SCN)$^-$ ligand, while the larger and softer Hg(II) ion occupies the B site. Thus, the compounds listed in Table 1 serve to demonstrate that the larger metal ion namely Hg(II) always binds to the soft S of the bridging thiocyanate ligand. In the compounds MnHg(SCN)$_4$ and MnHg(SeCN)$_4$, the Mn(II) ion occupies the A site and is bonded to N of thiocyanate. In addition, it has been observed in several structurally characterized examples of related thiocyanate compounds containing the {MnN$_4$O$_2$} octahedral unit (18-22), that the Mn(II) ion always binds at the N site of thiocyanate (Supplementary material, Fig. S1). Going by the preference of the A and B metal ions and the well documented affinity of Mn(II) to bind at the nitrogen site of thiocyanate in several bimetallic thiocyanates [6, 14, 18-22], structural chemistry principles rule out the substitution of Hg in CdHg(SCN)$_4$ by Mn to form **CdMn(SCN)$_4$** with Mn situated in the B site. However, the authors claimed that they have successfully grown crystals of CdMn(SCN)$_4$. A perusal of the details of growth and the characterization of this 'so called' CdMn(SCN)$_4$ crystal in the title paper reveals that the authors' claim is dubious as shown below.

*Growth of a 'so called' CdMn(SCN)$_4$ crystal*

The authors reported that by taking commercially available AR grade reagents namely KSCN, CdCl$_2$ and MnCl$_2$ in a 1:1:1 mole ratio in water, crystals formulated as CdMn(SCN)$_4$ can be grown as per their proposed reaction scheme in equation 1.

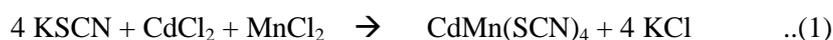

4 KSCN + CdCl$_2$ + MnCl$_2$  →  CdMn(SCN)$_4$ + 4 KCl          ..(1)

The reason for the disparity in the mole ratio (1:1:1) of reagents taken by authors, in contrast to equation 1, which requires a 4:1:1 reagent mole ratio, is not explained. Since analytical grade reagents

4(all are freely water soluble) were used for crystal growth, the statement of the authors, '*undissolved materials were filtered out and the clear solution was left undisturbed to slow evaporation of the solution and gradual growth*' clearly indicates the questionable nature of the entire synthetic protocol.

*A tetragonal crystal with unequal a and b unit cell parameters*

Based on the unit cell parameters which the authors claim to have determined, they reported that the crystals obtained from the reaction in equation 1 belong to the tetragonal crystal system and under the heading single crystal XRD the authors stated, '*The observed results indicate that the crystal belongs to tetragonal crystal system and the determined unit cell parameters are a = 12.0835 Å, b = 12.1349 Å, c = 8.5574 Å and α = 89.7128, β = 90.053, γ = 90.0023 and volume = 1254 Å$^3$*'. A tetragonal crystal system is characterized by the unit cell relation *a = b ≠* c; α = β = γ = 90°, which is described in all standard text books [23]. Hence the reported cell in the title paper appears to redefine the tetragonal system and the assignment of tetragonal system not only contradicts all known laws of crystal physics but is also not in accordance with the definition of a tetragonal crystal system in the International Tables for Crystallography [24]. In order to justify such an incorrect assignment, the authors stated, '*The family of crystals such as cadmium mercury thiocyanate (CMTC), ferrous mercury thiocyanate (FMTC), zinc mercury thiocyanate (ZMTC) and zinc cadmium thiocyanate (ZCTC) also belong to tetragonal crystallographic system*'. The reported single crystal study is dubious as shown below.

**Table 2** Unit cell data of AB(SCN)$_4$ compounds crystallizing in *I$\bar{4}$* space group

| No | Compound | *a* (Å) | *b* (Å) | *c* (Å) | V (Å$^3$) | Ref |
|---|---|---|---|---|---|---|
| 1 | CoHg(SCN)$_4$ | 11.109 | 11.109 | 4.379 | 540.41 | 5 |
| 2 | ZnHg(SCN)$_4$ | 11.0912(4) | 11.0912(4) | 4.4414(4) | 546.36(6) | 7 |
| 3 | MnHg(SCN)$_4$ | 11.324(3) | 11.324(3) | 4.270(2) | 547.5(3) | 6 |
| 4 | CdHg(SCN)$_4$ | 11.487(3) | 11.487(3) | 4.218(1) | 556.6(2) | 8 |
| 5 | ZnCd(SCN)$_4$ | 11.135(2) | 11.135(2) | 4.3760(10) | 542.6(2) | 9 |
| 6 | ZnCd(SeCN)$_4$ | 11.3420(1) | 11.3420(1) | 4.6326(1) | 595.942(15) | 12 |
| 7 | CoHg(SeCN)$_4$ | 11.281(3) | 11.281(3) | 4.6207(12) | 588.0(3) | 13 |
| 8 | MnHg(SeCN)$_4$ | 11.4545(4) | 11.4545(4) | 4.5803(4) | 600.96(6) | 14 |
| 9 | CdHg(SeCN)$_4$ | 11.6579(7) | 11.6579(7) | 4.5109(4) | 613.06(8) | 12 |
| 10 | CdMn(SCN)$_4$ | **12.0835** | **12.1349** | **8.5574** | **1254** | 15 |

The odd values in bold are reported in the commented paper

A comparison of the unit cell parameters of reported AB(XCN)$_4$ (X = S or Se) compounds (Table 2) all of which crystallize in the tetragonal space group $I\bar{4}$ , clearly show that in these compounds, the cell volumes depend on the size of A, B, S or Se. With increasing size of A ion (0.72 Å for Co$^{+2}$ to 0.92 Å for Cd$^{+2}$) the cell volume increases (see entry Nos 1 to 4). Replacement of S by Se results in a pronounced increase in cell volume. Substitution of Hg in ZnHg(SCN)$_4$ (entry No. 2) by Cd in ZnCd(SCN)$_4$ (entry No. 5) results in a slight decrease in cell volume. Going by this trend, the expected cell volume of for a compound having the composition CdMn(SCN)$_4$ should be less than that for the known compound CdHg(SCN)$_4$ (556.6 Å$^3$) if such a compound indeed crystallizes in the tetragonal system. In such a compound the Mn will occupy the A site. It is to be noted that the replacement of a larger ion like Hg(II) (ionic radius 1.10 Å) by Mn(II) (ionic radius 0.80 Å) can never result in more than double of the unit cell volume as has been claimed by the authors. The only set of data in table 2 which does not follow any trend and also look very odd / unusual are the unit cell values reported in the title paper, making one wonder if these parameters were actually determined. It is also not clear as to why the authors did not determine the structure of their crystal which is normally done in any single crystal work and substantiate the same in the form of a cif file. One finds it strange that molecular formula of a crystal is proposed based only on unit cell data.

*Intense signal at ~3450 cm$^{-1}$ in the IR spectrum*

The infrared spectrum in the title paper gives a clear indication that the proposed formula for the so called CdHg(SCN)$_4$ is inappropriate. The spectrum exhibits the most intense signal at ~3450 cm$^{-1}$ (O-H stretching vibration) and another at ~1620 cm$^{-1}$ (O-H bending vibration) which is never expected to be observed for any AB(SCN)$_4$ type compound, since it has no oxygen atom in its formula. The signal at ~3450 cm$^{-1}$ is so intense that it dwarfs all other signals clearly indicating that this is a characteristic signal of the crystal under study and not due to atmospheric moisture or due to absorption of moisture by KBr used for recording spectrum. The authors did neither list this intense band nor made a mention of it in their discussion but confidently concluded that the presence of functional groups was confirmed by IR technique. Given the reaction conditions and reagents used for crystal growth, the intense signal at ~3450 cm$^{-1}$ can only be assigned for a O-H stretching vibration. Interestingly the authors attributed the first weight loss in the TG-DTA thermogram for loss of water embedded within the crystal. It is not clear as to how water embedded in a crystal cannot form a part of the molecular formula of the crystal. The IR characterization would have been meaningful for the proposed formula CdMn(SCN)$_4$ had the authors identified and assigned the Cd-N and Mn-S vibrations based on literature precedence [25, 26]. Greenwood and Earnshaw [27] have reported that IR spectrum provides only a preliminary indication of the coordination mode of thiocyanate ion and IR is not a reliable criterion for product characterization in view of the mixing of the $\upsilon$(CN) and $\upsilon$(CS) group vibrations, and have instead suggested X-ray diffraction studies as the most reliable data for crystalline materials. Although the true identity of the crystal described in the title paper cannot be determined, the intense signal at ~3450 cm$^{-1}$ in the IR spectrum together with the unusually large unit cell volume is more than sufficient to convincingly prove that the crystalline product claimed to have been prepared is certainly not CdMn(SCN)$_4$.

**Conclusions**

It is appalling to note that a crystalline solid can be formulated based only on unit cell data without determining its structure. The assignment of tetragonal crystal system for an unit cell with differing *a* and *b* parameters is a serious error and unacceptable. Many doubts arise from the reading of the paper, starting from the research aim of the authors to substitute Hg in CdHg(SCN)$_4$ by Mn for the growth of CdMn(SCN)$_4$ crystal, which is impractical. The paper is completely erroneous and full of unsubstantiated claims and the crystal growth details and interpretation of all experimental data are questionable. It is very unfortunate that such a report was published after a peer-review process. The importance of scientific publications calls for a more careful scrutiny of submitted papers especially the ones claiming growth of novel crystals, considering the frequent reporting of improperly characterized compounds.